\documentclass[aip,graphicx,journal=jcp]{revtex4-1}

\draft 

\usepackage[version=3]{mhchem} 
\usepackage{dcolumn}
\usepackage{mathrsfs}
\usepackage[loose,hang]{subfigure}
\usepackage[usenames,dvipsnames]{color}
\usepackage{graphicx}
\usepackage{tikz}
\usepackage{amssymb,amsmath}
\usepackage{txfonts}
\usepackage{ulem}
\usepackage{epstopdf}
\DeclareFontFamily{OT1}{pzc}{}
\DeclareFontShape{OT1}{pzc}{m}{it}{<-> s * [1.10] pzcmi7t}{}
\DeclareMathAlphabet{\mathpzc}{OT1}{pzc}{m}{it}

\newcounter{tempEquationCounter}
\newcounter{thisEquationNumber}

\begin{document}

\def\ket#1{ $ \left\vert  #1   \right\rangle $ }
\def\ketm#1{  \left\vert  #1   \right\rangle   }
\def\bra#1{ $ \left\langle  #1   \right\vert $ }
\def\bram#1{  \left\langle  #1   \right\vert   }
\def\spr#1#2{ $ \left\langle #1 \left\vert \right. #2 \right
\rangle $ }
\def\sprm#1#2{  \left\langle #1 \left\vert \right. #2 \right
\rangle   }
\def\me#1#2#3{ $ \left\langle #1 \left\vert  #2 \right\vert #3 
\right\rangle $ }
\def\mem#1#2#3{  \left\langle #1 \left\vert  #2 \right\vert #3 
\right\rangle   }
\def\redme#1#2#3{ $ \left\langle #1 \left\Vert
                  #2 \right\Vert #3 \right\rangle $ }
\def\redmem#1#2#3{  \left\langle #1 \left\Vert
                  #2 \right\Vert #3 \right\rangle   }
\def\threej#1#2#3#4#5#6{ $ \left( \matrix{ #1 & #2 & #3  \cr
                                           #4 & #5 & #6  } \right)$}
\def\threejm#1#2#3#4#5#6{  \left( \matrix{ #1 & #2 & #3  \cr
                                           #4 & #5 & #6  } \right) }
\def\sixj#1#2#3#4#5#6{ $ \left\{ \matrix{ #1 & #2 & #3  \cr
                                          #4 & #5 & #6  } \right\}$}
\def\sixjm#1#2#3#4#5#6{  \left\{ \matrix{ #1 & #2 & #3  \cr
                                          #4 & #5 & #6  } \right\} }

\def\ninejm#1#2#3#4#5#6#7#8#9{  \left\{ \matrix{ #1 & #2 & #3  \cr
                                                 #4 & #5 & #6  \cr
                         #7 & #8 & #9  } \right\}   }
%
%
%
%
\newcommand{\pr}{^{\prime}}
\newcommand{\bfx}{\overrightarrow{x}}
\newcommand{\bfp}{{\bf p}}
\newcommand{\la}{\langle}
\newcommand{\ra}{\rangle}
\newcommand{\rp}{{p}}
\newcommand{\rpp}{{{p}^{\prime}}}
\newcommand{\rx}{{x}}
\newcommand{\eps}{{\mbox{$\epsilonup$}}}
\newcommand{\be}{\begin{eqnarray}}
\newcommand{\ee}{\end{eqnarray}}
\newcommand{\ba}{\begin{array}}
\newcommand{\ea}{\end{array}}
\newcommand{\proc}{{alpha decay }}
\newcommand{\balpha}{{\mbox{\boldmath$\alpha$}}}
\newcommand{\bsigma}{{\mbox{\boldmath$\sigma$}}}
\newcommand{\bnabla}{{\mbox{\boldmath$\nabla$}}}
\newcommand{\bmu}{{\mbox{\boldmath$\mu$}}}
\newcommand{\bfr}{{\bf r}}
\newcommand{\bfb}{{\bf b}}
\newcommand{\bfe}{{\bf e}}
\newcommand{\bfk}{{\bf k}}
\newcommand{\bfK}{{\bf K}}
\newcommand{\bfR}{{\bf R}}
\newcommand{\bfA}{{\bf A}}
\newcommand{\der}{\mathrm{d}}
\newcommand{\bbeta}{{\mbox{\boldmath$\beta$}}}
\newcommand{\bomega}{{\mbox{\boldmath$\omega$}}}
\newcommand{\tikzcircle}[2][red,fill=red]{\tikz[baseline=-0.5ex]
\draw[#1,radius=#2] (0,0) circle ;}
\newcolumntype{.}{D{.}{.}{-1}}
\newcolumntype{d}[1]{D{.}{.}{#1}}
\renewcommand{\thefootnote}{\alph{footnote}}
\renewcommand{\emph}[1]{\textit{#1}}
\definecolor{aqua}{RGB}{0,255,255}


\title{Valence-bond Non-equilibrium Solvation Model for a Twisting Monomethine Cyanine} 



\author{Sean McConnell}
\author{Ross H. McKenzie}
\author{Seth Olsen}
\email{seth.olsen@uq.edu.au}
\affiliation{School of Mathematics and Physics, The University 
of Queensland, Brisbane 4072, Australia}


\date{\today}

\begin{abstract}
	We propose and analyze a two-state valence-bond model of non-equilibrium solvation effects on the excited-state twisting reaction of monomethine cyanines.  Suppression of this reaction is thought responsible for environment-dependent fluorescence yield enhancement in these dyes.  Fluorescence is quenched because twisting is accompanied via the formation of dark twisted intramolecular charge-transfer (TICT) states.  For monomethine cyanines, where the ground state is a superposition of structures with different bond and charge localization, there are two possible twisting pathways with different charge localization in the excited state.  For parameters corresponding to symmetric monomethines, the model predicts two low-energy twisting channels on the excited-state surface that lead to a manifold of twisted intramolecular charge-transfer (TICT) states. For typical monomethines, twisting on the excited state surface will occur with a small barrier or no barrier. Changes in the solvation configuration can differentially stabilize TICT states in channels corresponding to different bonds, and that the position of a conical intersection between adiabatic states moves in response to solvation to stabilize either one channel or the other.  There is a conical intersection seam that grows along the bottom of the excited-state potential with increasing solvent polarity.  For monomethine cyanines with modest-sized terminal groups in moderately polar solution, the bottom of the excited-state potential surface is completely spanned by a conical intersection seam. 
\end{abstract}


\maketitle 

\section{Introduction}
\begin{figure}
\centering
\includegraphics[width=10cm]{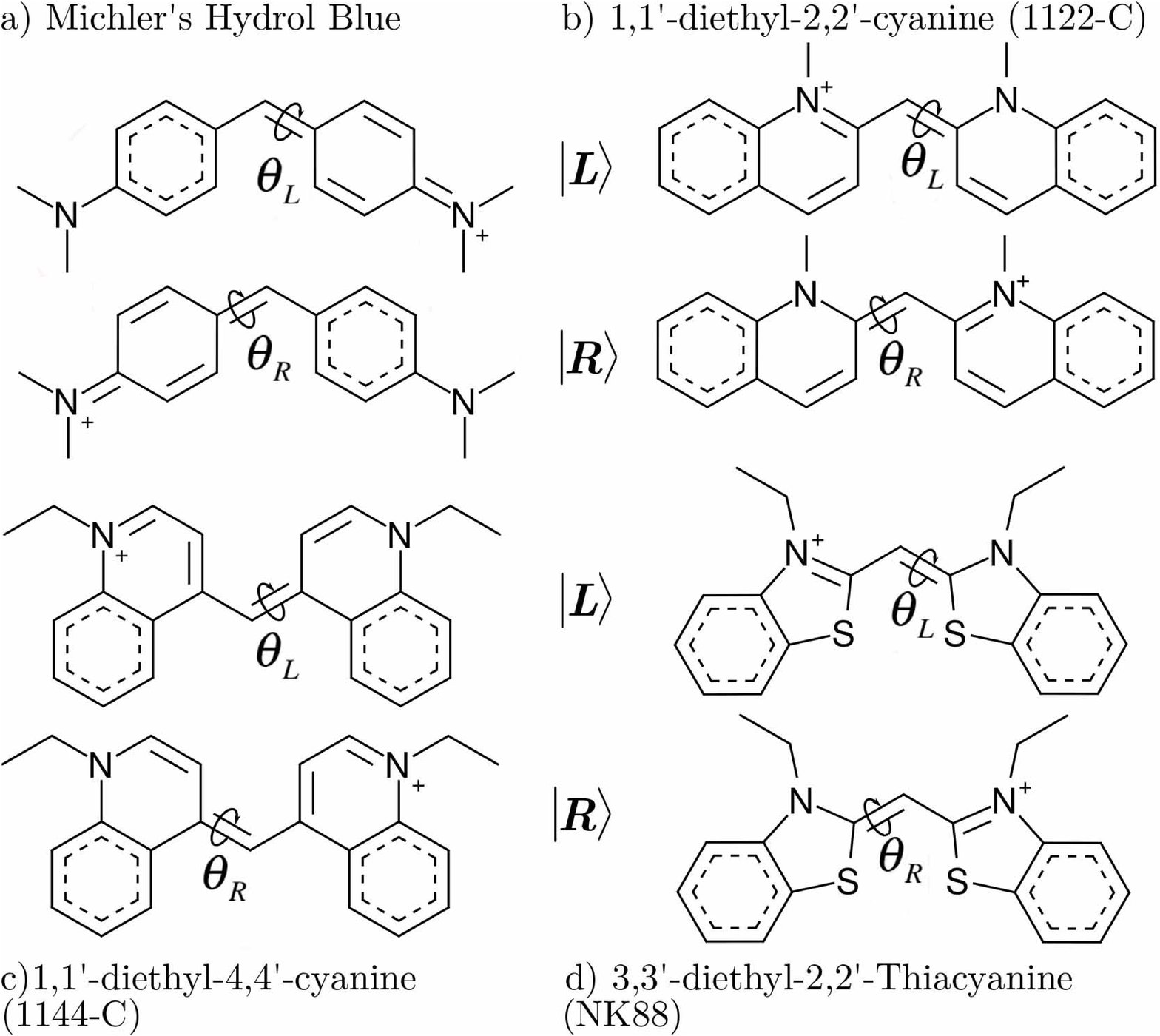}
\caption{\label{fig:Scheme} Some symmetric monomethine cyanine dyes. The ground 
state of these cyanines is an even superposition of states $\left|R\right
\rangle$ and $\left|L\right\rangle$ with different charge and bond localization.  In Michler's hydrol blue (upper left) the physical charge carrier is an electron, but a hole in other dyes shown, explaining the different disposition of bond and formal charge.}
\end{figure}

	Monomethine cyanines have a distinguished role in the history of chemical science and industry.\cite{Rose1926,Brooker1942}  The description of their optical properties was an early target of quantum electronic structure models.\cite{Platt1956,Platt1961}  The optical properties of monomethine dyes and push-pull polyenes can be described by two-diabatic-state models based on the concept of resonance between configurations with opposing charge and bond localization.\cite{Platt1956,Lu1994,Thompson1998,Barzoukas1996,Painelli1999,Boldrini2002}.  The resonating diabatic states are shown for four typical symmetric monomethine dyes in \ref{fig:Scheme}.

	A distinguishing feature of monomethine cyanines is the environmental sensitivity of their fluorescence emission yield.\cite{Benson1977}  The dyes are practically non-fluorescent in fluid solutions, but can become highly fluorescent when bound to biomolecules or other constraining media.\cite{Hutten1945,Baldridge2011a,Baldridge2011b,Paige2011,Brey1977,Clark1986,Murphy1994}  This underlies the use of monomethine  cyanines as fluorescent turn-on labels in biological imaging.  The fluorescence enhancement is interpreted as suppression of a bond-twisting process in the excited state, which would otherwise lead to dark twisted intramolecular (TICT) states\cite{Rettig1986}. The TICT states in monomethine cyanines are dark, so population of these states is associated with ultrafast loss of transient fluorescence.  The TICT are characterized by a small adiabatic gap, and nonadiabatic decay can occur by conical intersections that occur at twisted geometries.\cite{Olsen2012a}

	Enhancement of fluorescence due to suppression of excited-state twisting motion is the defining characteristic of a "molecular rotor"\cite{Amdursky2012,Haidekker2010}.  This behavior is not limited to monomethine cyanines.  An example of a non-methine molecular rotor is the amyloid fibril-sensing dye Thioflavin-T\citep{Amdursky2012}.  This molecule undergoes twisting in the excited state, allowing ultrafast internal conversion via twisted intramolecular charge-transfer (TICT) states with small or vanishing electronic gap\cite{Rettig1986}.  The TICT states in Thioflavin-T and Auramine-O are dark, so that the twisting reaction is accompanied by a fast transient fluorescence decay and negligible steady-state quantum yield.  The fluorescence is enhanced upon binding to biomolecular environments or in viscous solution. The nonradiative decay rates of Thioflavin-T and the monomethine dye Auramine-O have been extensively characterized by Huppert and coworkers in a series of glass-forming alcohols, under conditions spanning 10 orders of magnitude in the viscosity\cite{Erez2012d}.   These studies supported earlier indications\cite{VanderMeer2000} that the nonradiative decay time could be estimated from the damped rotational Stokes drift of a spherical rotor down a linear potential through an angle of one radian. The reaction coordinate is taken to be rotation of a double bond to form a dark TICT state.  A useful estimate of the rotational timescale (which describes the non-radiative decay timsecale) was found to be
\begin{equation}\label{eq:trot}
\tau_{rot}=\frac{2 \eta V}{torque}
\end{equation},
where $\eta$ is the medium viscosity, $V$ is the effective volume of the rotor (20 $\r A^3$ were taken for Thioflavin-T or Auramine-O), and $torque$ is the mean torque on the rotor over a rotation of one radian. A key point was that the non-radiative decay timescale was found to depend linearly on the medium viscosity.  A linear viscosity dependence of the nonradiative decay rate was also observed in experiments on the triphenylmethane dye Crystal violet, which decays via a similar mechanism.\cite{BenAmotz1987a,BenAmotz1987b,BenAmotz1987c}  
	
	Different possible situations for a one-dimensional molecular rotor system with dark TICT states are illustrated schematically in Figure \ref{fig:TICTOutcomes}.  For twisting that is both favorable in the excited state and without barrier, one expects ultrfast (~1ps) decay of fluorescence as the TICT state is formed from the bright Franck-Condon (FC) state.  Provided depletion of TICT excited state is not rate-limiting, which our results support (see below), ground state recovery should track the fluorescence decay. For twisting that is favorable but activated, one expects slower fluorescence decay.  Again, if the depletion of the excited TICT states is not rate-limiting, the ground state recovery should follow the fluorescence decay.  When TICT state formation is not favorable, we expect that they do not form on a relevant timescale and the molecule is fluorescent.  If our assumption about fast depletion of the TICT excited-state were wrong, so that populuation accumulates on the excited state, then the same pattern would appear in the ultrafast fluorescence, but we would expect the ground state recovery to lengthen independently.

	For molecular rotors with the dimensions characteristic of monomethine cyanines, the timescale for generation of a charge-transfer state by twisting and the timescale for solvation of the same charge-transfer state cannot be reliably separated. The nonradiative decay times for Thioflavin-T and Auramine-O in glass-forming solvents scale with the transverse Debye dielectric relaxation time\cite{Erez2012d}, but are 1-2 orders of magnitude smaller\cite{Erez2012d}.  This is consistent with nonradiative decay times that are comparable to the Debye \textit{longitudinal} relaxation time.  The longitudinal relaxation time is the appropriate timescale to describe dielectric relaxation following a sudden change in the molecular charge density\cite{Nitzan2006sol}.  For alcoholic solutions, the longitudinal relaxation time will be ~10 times smaller than the transverse timescale\cite{Barthel1990b}, which is comparable to estimated twisting times.  Although solvent-dependent effects can be observed at sub-ps timescales for Auramine-O, these were assigned to a distinct dynamical origin.\cite{Erez2014}  Heisler and colleagues showed for Auramine-O in nanoconfined water droplets that restriction of solvation by the confinement slowed the fluorescence upconversion decay.\cite{Heisler2009}  Results reported by Kondo and coworkers for Auramine-O in bulk solutions indicated that the dominance of solvation dynamics vs reaction coordinate friction in determining the fluorescence upconversion decay could not be determined for several solvents excepting water, for which solvation dynamics was dominant.\cite{Kondo2010}  The \textit{N,N'}-dimethylaniline rotors in Auramine-O are identical to those of Michler's Hydrol Blue in Figure \ref{fig:Scheme}, and so the timescales for damped twisting are expected also to be the same.  

	If the timescales for solvation and for twisting are comparable, then a faithful description of the physics must treat these processes at a comparable level of approximation.  A successful yet simple approach for the one-dimensional rotor \textit{N,N'}-dimethylbenzoaminonitrile (DMABN), which forms an emissive TICT state on a timescale much longer than common for monomethine cyanines, has been to develop two-dimensional models coupled to a twisting coordinate and a single solvation coordinate that gauges the polarization state of the medium.\cite{Fonseca1994a,Fonseca1994b,Kim1997,Nordio1997}

	The identification of monomethine cyanines as molecular rotors raises an interesting question, because the ground states of these molecules are a superposition of structures with different $\pi$-bond localization (c.f. Figure \ref{fig:Scheme}).  In a molecule where the double bond can be in multiple places, how does one choose the appropriate reaction coordinate to describe double bond twisting in the excited state?  Ab initio studies of the excited state surface in monomethine cyanines indicate that there are not one but \textit{two} low-energy twisting channels on the excited-state potential energy surface\cite{Improta2005}.    

	General models of TICT electronic structure suggest that as the bond twisting proceeds, the lowest two adiabatic states should become polarized and form a twisted intramolecular charge-transfer (TICT) system.\cite{Rettig1986,Dekhtyar1999,Olsen2009,Olsen2012a}.  This picture is upheld by more detailed quantum chemical models\cite{Improta2005,Sanchez2000}. However, twisting one or the other bond will destabilize the diabatic states by different amounts -- because the $\pi$-bonds are in different places.  Accordingly, the polarity of the TICT adiabatic states will depend upon the bond that is twisted\cite{Olsen2012a,Olsen2009}.  The dependence of the TICT polarity on the identity of the twisted bond can also be explained using simple molecular orbital theories.\cite{Dekhtyar1999}  The correlation between the path taken and the TICT polarity suggests that a polar environment can distinguish the twisting pathways, and could influence their relative yields.  This has been suggested as the mechanism that underlies reversible photochromism and fluorescence switching in fluorescent proteins\cite{Olsen2010}.  Fluctuation in the charge-coordination environment have been predicted to influence the bifurcation and availability of twisting pathways in photoactive yellow protein chromophores\cite{Gromov2011,Ko2008,Knoch2014}.

	Modern theoretical photochemistry suggests the location, topography and energetics of conical intersection seams between adiabatic electronic states are important for understanding the mechanism and yield of photochemical reactions\cite{Bernardi1996,Klessinger1995,Levine2007}.  Ab initio calculations and general models indicate that conical intersections can occur in monomethine dyes in configurations where one or more bonds is twisted, so that the $\pi$ orbitals on the fragments across the bond overlap weakly\cite{Improta2005,Olsen2009,Olsen2012a}.  The non-adiabatic coupling (to nuclear displacement) diverges at conical intersections, so that internal conversion is expected to be dominated by configurations near these points.  Unlike transition states that separate different basins on a given surface, conical intersections are multidimensional seams embedded in the configuration space of the molecule\cite{Levine2007}.  Although conical intersections can sometimes be discussed in terms of their minimum-energy configurations, there are cases where the relevant nonadiabatic dynamics is thought to occur far removed from the minimum-energy regions of the intersection seam\cite{Virshup2012,Coe2008}.  Since the twisted conical intersections in monomethine dyes arise from intersection of diabatic states with different charge localization, changes in the solvent configuration could very plausibly affect the positioning, topology and/or energetics of the seam.  

	The effect of polar solvation on the structure of conical intersection seams in a retinal protonated Schiff base model has been studied in a series of papers by Hynes and coworkers\cite{Burghardt2004,Burghardt2006,Spezia2006,Malhado2011}.  In these papers, it was shown that solvent effects determined the placement and energy of conical intersection seams, as well as the coordinates that accessed them.  A tendency for charge-transfer conical intersections to be stabilized by solvents excited-state potential in solvated systems has been observed in excited-state dynamics simulations\cite{Toniolo2003,Virshup2009}.  
	
	In this paper, we seek to develop a better intuition for how non-equilibrium solvation affects the excited-state twisting process in monomethine cyanines, with special attention to the issue that there are two possible twisting pathways that can be distinguished by the solvent. 

	One reason why it is important to consider simple models is that simple models make it easy to keep track of how the ideas that go into the model relate to the predictions that come out.  That way, a challenge to the model can be framed as a challenge to the underlying ideas, rather than to some detail of the model parameterization.  In this paper, we focus on only a few ideas that are simple extensions of the old notion of cyanine electronic structure as a resonance between structures with opposing charge- and bond-order localization (cf. Figure \ref{fig:Scheme}).  

	Our model describes favorable excited-state twisting with no or a small barrier for typical monomethine cyanines in solution, leading to a manifold of twisted intramolecular charge-transfer (TICT) configurations with small or no electronic gap.  The driving force for twisting in the isolated dye is small for dyes with absorption in the red, and increases with increasing absorption energy for a given $\pi$-bond stabilization energy.  The model predicts that solvation of products along one twisting pathway tends to destabilize the alternate pathway, so that the model can describe solvent-driven selection of distinguishable TICT channels.  The position and energy of a conical intersection in the TICT manifold is shown to change with the solvent configuration.  The fraction of TICT states that can be made to coincide with the intersection seam grows with the solvent reorganization energy.  For couplings corresponding to typical monomethine cyanines in  polar solvents, there is no minimum or "basin" that can be associated with an adiabatically excited TICT system.  Instead, the bottom of the excited state potential coincides with a conical intersection for monomethine cyanines in even moderately polar solvents. 

	Here is how the paper is organized.  In Section 2, we introduce the model, describing its physical content and parameterization.  Section 3 discusses the potential energy surfaces (PESs) given by the model.  We show that the PESs of the model can be qualitatively categorized according to a phase diagram, and describe relevant features of the phase diagram.  Section 4 discusses results in the context of experiments on the nonradiative decay of monomethines.  Section 5 concludes.

\section{Model Hamiltonian}
	The model Hilbert space is two-dimensional, and spanned by diabatic basis states distinguished by charge and bridge $\pi$-bond order as in Figure \ref{fig:Scheme}.  In this basis, the Hamiltonian is written as (\ref{eq:Ham_mat}).
\begin{equation}
\hat{\mathbf{H}}=\hat{\mathbf{H}}_{Dye}+\hat{\mathbf{H}}_{Sol}=
\underbrace{\left(
	\begin{array}{cc}
J\sin^2{\theta_L}&-\beta\cos{\theta_L}
\cos{\theta_R}\\
-\beta\cos{\theta_L}\cos{\theta_R}&J
\sin^2{\theta_R}
	\end{array}
\right)}_{\hat{\mathbf{H}}_{Dye}}+\underbrace{\lambda\left(
	\begin{array}{cc}
\left(\frac{s+1}{2}\right)^2&0\\
0&\left(\frac{s-1}{2}\right)^2
	\end{array}
\right)+\left(\frac{1}{\sqrt{n}}-1\right)\frac{1}{2 r}}_{\hat{
\mathbf{H}}_{Sol}}.\label{eq:Ham_mat}
\end{equation}

	The Hamiltonian is comprised of two subcomponents, a dye Hamiltonian 
$\hat{\mathbf{H}}_{Dye}$, which is a function of the bridge-bond twist angles $\theta_L$ and $\theta_R$, and a solvent interaction Hamiltonian $\hat{\mathbf{H}}_{Sol}$, which is a function of a single solvent configuration coordinate $s$, which interpolates between equilibria with $\lvert L\rangle$ $(s=-1)$ and $\lvert R\rangle$ $(s=1)$. The parameters $\beta$, $J$ and $\lambda$ are the electronic coupling strength, the $\pi$-bond stabilization energy, and the diabatic solvent reorganization energy.  The solvent interaction term $\hat{\mathbf{H}}_{Sol}$ includes a configuration-independent scalar solvation energy that 
depends on the refractive index $n$ of the medium and the Born cavity radius $r$ characterizing the relevant terminal heterocycle.  This term is the same for both diabatic states in a symmetric dye and does not vary with the configuration of the dye or solvent; it can be removed by resetting the zero of energy. 

	The transformation between the diabatic and adiabatic bases can be 
parameterised as a rotation by an angle $\phi$
\begin{equation}	\label{eq:GMH_diabat_def}
	\left[
	\begin{array}{c}
	|0\rangle\\	
	|1\rangle
	\end{array}
	\right]=
	\left[
	\begin{array}{cc}
	\cos \phi & \sin \phi \\	
	-\sin \phi & \cos \phi
	\end{array}
	\right]
	\left[
	\begin{array}{c}
	|L\rangle\\
	|R\rangle
	\end{array}
	\right].
\end{equation}	After resetting the zero of 
energy to eliminate the scalar solvation energy, the 
eigenvalues of the Hamiltonian (\ref{eq:Ham_mat}) are 
\begin{equation}\label{eq:evals}
E_0=\bar{E}-\frac{\Delta\mathpzc{E}}{2},\nonumber\\
E_1=\bar{E}+\frac{\Delta\mathpzc{E}}{2},\nonumber\\
\end{equation}
where
\begin{equation}
\bar{E}=\frac{1}{2}\left(\frac{\lambda}{2}\left(s^2+1\right)
 + 
 J\left(\sin^2{\theta_L} + \sin^2{\theta_R}\right)\right), 
\nonumber\\
\Delta \mathpzc{E}=\sqrt{\left(2\beta\cos{\theta_L}\cos{\theta_R}\right)^2 
+ \left(J\left(\sin^2{\theta_L}-
\sin^2{\theta_R}\right)+\lambda s\right)^2}.
\end{equation}

\subsection{Dye Hamiltonian}
	The diabatic energies vary with the twisting of the bridge bonds, through the angles $\theta_L$ and $\theta_R$.  The dependence reflects distinguishable $\pi$-bond localization in the diabatic states in Figure \ref{fig:Scheme}.   Each diabatic state has $\pi$-bonding character localized on a distinct bridge bond.  The $\pi$-bond stabilization energy $J$ 
is the energy required to twist a $\pi$ -bond to $\theta = \frac{\pi}{2}$, breaking the bond.

\begin{align}
\label{eq:ham_mat_elems}
\mem{L}{\hat{\mathbf{H}}_{Dye}}{L}=J\sin^2{\theta_L}\\
\mem{R}{\hat{\mathbf{H}}_{Dye}}{R}=J\sin^2{\theta_R}.
\end{align}

	We are concerned with the interaction of twisting and solvation displacements, because the corresponding timescales in monomethine cyanines appear to be non-separable.  A literal interpretation of the valence-bond structures in Figure \ref{fig:Scheme} implies also strong coupling to bond-stretching displacements.  Such displacements are important for understanding linear and nonlinear spectra of these and similar systems\cite{Gorman1993,Sissa2012,Campo2010,Lu1994,Thompson1998,Thompson1999}.  

	The off--diagonal coupling has one term representing charge transfer 
between rings via sequential transfers across both bridge bonds. The direct transfer element between the ring is taken to be neglible, so that the lowest order couplings between the (many-body) diabatic states go as the product of transfer terms for each bond, and the modulation of transfer across each bond goes as the overlap between $\pi$-orbitals on the bonding fragments. Similar strategies have been applied to the nonlinear solvation of systems with activated TICT formation along a single twisting coordinate coupled to a solvation mode.

\begin{equation}\label{eq:ham_mat_od}
\mem{L}{\hat{\mathbf{H}}_{Dye}}{R}=-\beta
\cos{\theta_L}\cos{\theta_R}.
\end{equation}

	The parameter $\beta$ determines the magnitude of the electronic coupling
between diabatic states. It is equal to the adiabatic gap at the (planar) ground state geometry.
The form of equations (\ref{eq:ham_mat_elems}) and 
(\ref{eq:ham_mat_od}) can be justified using valence bond 
theory\cite{Tennyson1981,Fox1985,Bernardi1988,Shaik2004}, in 
so far as the single-particle transfer element goes as the overlap between 
atomic \textit{p} orbitals.  The parameterization assumes a well-defined nodal plane associated with an $sp^2$-hybridized monomethine bridge, so that the twist angles $(\theta_l,\theta_r)$ can be unambiguously defined.  

\subsection{Parameterisation of the dye Hamiltonian}
	The electronic coupling parameter $\beta$ is equal to the electronic gap when the dye is planar and the solvent is unpolarized, which is true at the ground state minimum.  Since monomethine dyes have strong, narrow absorbance lines this parameter can be estimated from absorbance spectroscopy as the energy corresponding to the band maximum.  Alternatively, it can be estimated using quantum chemistry calculations.
\begin{equation}
\Delta \mathpzc{E}(0,0)=2\beta
\end{equation}
The absorbance of monomethine cyanines is usually sharp, and solvatochromism is negligble\cite{Griffiths1976}.  For dyes such as in Figure \ref{fig:Scheme}, with absorbance in the visible, an appropriate value for $\beta$ would be in the range 0.8-1.7 eV.

	The $\pi$-bond energy $J$ is equal to the adiabatic gap if one bond is twisted and the other is planar (in an unpolarized solvent). 
\begin{equation}\label{eq:evals_t}
\Delta \mathpzc{E}\left(\pm\pi/2,0\right)=
\Delta \mathpzc{E}\left(0,\pm\pi/2\right)=
J .
\end{equation}
	This suggests the parameter $J$ is in-principle estimable from experiments, if transient fluorescence from the excited TICT state could be collected in sufficiently non-polar solvents.  Unfortunately, the transition dipole strength also vanishes at TICT states, making spectroscopic estimation difficult or unfeasible in most cases.  In such cases, it may be easiest to resort to calculation of the gap using quantum chemistry computations at a suitable model of the twisted state geometry.

	General considerations suggest that the parameter $J$ should be in the range 1.0-2.0 eV.  This estimate comes from the identification of $J$ with the energy associated with twisting a $\pi$-bond in a diabatic state with clearly defined bond alternation.  To the extent that the energy of a localized double bond is transferrable, we expect that $J$ should be at most comparable to the ground-state twisting energy for a molecule with definite single-double bond alternation in the adiabatic ground state.  The reason for the "at most" in the previous sentence is that the initial state for bond-twisting in a molecule with definite ground state bond alternation is near the minimum for the relevant diabatic state, this will not be the case in a symmetric monomethine where the ground state geometry is ambiguous.  A reasonable reference system of similar size to the molecules in Figure \ref{fig:Scheme} might be an asymmetrically substituted stilbene.  The singlet ground state barriers for stilbene and 4-styrylpyridinium have been reported as 1.78 eV and 1.69 eV, respectively\cite{Bortolus1970}.  Another route to a similar conclusion would be to note that the $\pi$-bond stabilization energy in H\"{u}ckel theory is just the resonance integral $\beta$, which is usually found to be in the range 1-2eV\cite{Cotton}.  In any case, we are led to the conclusion that an appropriate value for the barrier height will be ~1 eV.
	
	For some systems, calculations of the diabatic barrier height are available, as calculations of the adiabatic gap at a suitable model TICT geometry.  For near-resonant models of the GFP chromophore, which are not strictly symmetric but electronically almost so, this yields a barrier height of 0.9-1.0eV\cite{Olsen2012a}.  For the case of Michler's Hydrol Blue, we have an unpublished value of the twisted gap calculated using multi-state and multi-reference perturbation theory that is 1.0 eV.  For a slightly truncated model of the monomethine cyanine NK88 (cf. Figure \ref{fig:Scheme}), Santoro and coworkers have calculated twisted gaps in the range 1.3-1.5 eV using different electronic structure approaches\cite{Improta2005}.   These values for the diabatic barrier height $\beta$ are consistent with the arguments  above.
	
\subsection{Dye--solvent interaction}
	The interaction with the solvation field is described using a model of charge transfer between conducting spheres in a linear dielectric, with cavity radii and intercavity distance chosen to reflect the geometry of a typical monomethine with modest-sized rings.  This model is well-established in the electron transfer literature \cite{Nitzan2006marcus}.  It has also been successfully applied to model the anomalous solvatochromism of charged push-pull polyenes in solution.\cite{Laage2003a} The diabatic solvent reorganization energy $\lambda$ is calculated as the reorganization energy of charge transfer between the spheres.  Charged push-pull polyenes are chemically similar to monomethine dyes (cf. Figure \ref{fig:Scheme}) and are, essentially, strongly asymmetric polymethine cyanines.  

	The configuration of the solvent is parameterised by a single solvent 
coordinate $s$. Those configurations which are in 
equilibrium with the diabatic states $|L\rangle$ and $|R\rangle$ 
(c.f. Figure \ref{fig:Scheme}) we denote by $s=1$ and 
$s=-1$ respectively. The free energies of the diabatic 
states are shown in Figure \ref{fig:Nitzan}.

	The solvation interaction $\hat{\mathbf{H}}_S$ is diagonal in the diabatic basis.  The relevant matrix elements are
\begin{align}
\label{eq:Marcus_energies}
\left\langle L \left|\hat{\mathbf{H}}_S\right| L 
\right\rangle=G_0
+\lambda\left(\frac{s+1}{2}
\right)^2\\
\left\langle R \left|\hat{\mathbf{H}}_S\right| R 
\right\rangle=G_0
+\lambda\left(\frac{s-1}{2}
\right)^2,
\end{align}
where $\lambda$ is the dielectric reorganization energy and $G$ is the Born solvation energy of a charged conducting sphere of radius $r$ embedded in a linear dielectric (in CGS units)
\begin{equation}
\label{eq:solv_elems1b}
G_0=\left(\frac{1}{\epsilon}-1 \right)\frac{e^2}{2 r}.
\end{equation}
	For a symmetric monomethine, the cavity radii will be the same, and the solvation energy terms contribute only to the trace of the Hamiltonian.  They can be removed by resetting the zero of 
energy to $G_0$. 

	The physical interpretation of the solvent reorganisation energy 
$\lambda$ is the energy released by the non--adiabatic relaxation of 
the system to the minimum energy configuration of $|R\rangle$ at $s=1$ following excitation from $|L\rangle$ 
at $s=-1$, the expression for this is\cite{Nitzan2006marcus,Laage2003a}
\begin{equation}\label{eq:reorg_e}
\lambda\equiv\lambda(r,R)=\left(\frac{1}{\eps_\infty}-
\frac{1}{\eps}\right)e^2\left(\frac{1}{r}-\frac{1}{R}\right),
\end{equation}
where $r$ is the cavity radius, $R$ is the distance between 
cavity centers, and $\eps$ and $\eps_\infty$ are the static and high-frequency dielectric constants of the medium.  We have assumed the cavity radii are the same and that the transferred charge is equivalent to the fundamental charge $e$.

	For symmetric monomethine dyes such as shown in Figure \ref{fig:Scheme},
we expect $R\gtrapprox 2r$ and cavity radii 
$r\sim 2-4\r A$. Table \ref{tab:solv_reorg_en} shows the 
reorganisation energies at these values of $r$ and $R$. Values of 
$r\to 2\r A$ are representative of smaller endgroups, such as in dyes used as models of barrierless reaction kinetics\cite{Rettig1982,Akesson1986}.

\begin{table*}
\caption{\label{tab:solv_reorg_en} Reorganisation 
energies (in eV) representative of monomethine cyanines in some typical solvents. The values of the cavity 
size and endgroup separation $(r,R)$ are shown in brackets.  The ranges of r and R are chosen to span realistic ranges for monomethine cyanines with endgroups containing 1-2 rings with small substituents.}
	\begin{tabular}{|l|l|l|cc|cc|cc|}
	\hline
	Compound&\eps& \eps$_\infty$&
$(4,8)$&$(4,9)$&$(3,6)$&$(3,7)$&$(2,4)$&$(2,5)$\\
	\hline
	water 			&78.4&1.9&1.0&1.1&1.3&1.5&2.0&2.4\\
	acetonitrile		&36.6&1.8&1.0&1.1&1.3&1.5&1.9&2.3\\
	methanol			&33.0&1.8&1.0&1.1&1.3&1.5&1.9&2.3\\
	dichloromethane	&8.9&2.0&0.7&0.8&0.9&1.1&1.4&1.7\\
	chlorobenzene	&5.7&2.3&0.5&0.5&0.6&0.7&0.9&1.1\\
	\hline
	\hline
	\end{tabular}
\end{table*}

	Solvent reorganisation energies for a conducting-sphere model
of monomethines with reasonably small rings (like Figure 
\ref{fig:Scheme}), as outlined in table \ref{tab:solv_reorg_en}, are 
in excess of 1eV, close to or larger than typical values of $J$, according to the last section. Huppert and coworkers have used an approximate volume of ~20$\r A^3$ to calculate the nonradiative decay rate of auramine-O and Thioflavin-T, corresponding to a sphere of radius ~2$\r A$.  This happens to be just under half of the experimentally measured transition dipole moment for Michler's hydrol blue\cite{Looney1954}.  The agreement in estimates of the rotor size is satisfying, since one is estimated from a hydrodynamic model of the non-radiative decay time\cite{Amdursky2011}, while the other is obtained from an absorbance spectroscopy experiment\cite{Looney1954}.

	We will now summarize the ideas that have gone into the model Hamiltonian Equation \ref{eq:Ham_mat}.  
	\begin{itemize}
\item The model is built on a Hilbert space spanned by structures with distinct charge and $\pi$-bond localization.  This notion of the low-energy electronic structure of cyanines has a history in chemical theory that goes back to the early part of the last century\cite{Brooker1942}\cite{Griffiths1976}.   
\item We have introduced the idea that $\pi$-bonds have barriers to rotation that are not present in single ($\sigma$) bonds.  This concept is not new, and is one of the most recognized basic facts of structural organic chemistry.  The functional form of the barrier to rotation that we have used can be derived from valence-bond theoretical arguments\cite{Shaik2004}.  The $\pi$-bond stabilization energy can be estimated to be in the range 1-2eV based on textbook arguments\cite{Cotton}.   
\item The inter-ring coupling is a product of transfers onto and off of the bridge.  This reflects the idea that the direct (i.e. single-particle) inter-ring charge transfer coupling is negligible, which goes back as far as Pauling\cite{Pauling1939}.  Since the direct coupling vanishes, the coupling goes as a product of transfers across the bridge bonds in lowest order.  The dependence on the variables $\theta_L$ and $\theta_R$ across each bond go as the overlap of local $\pi$ fragment orbitals across the respective bond.  A lucid early description of the many-particle origin of the large ($\beta~1-2$eV) coupling in cyanines was given by Moffitt\cite{Moffitt1950}.  Essential aspects of this analysis have recently been confirmed again using a quasi-diabatic analysis of the electronic structure of series of monomethine cyanines\cite{Olsen2012a}.
\item We have introduced the notion that the solvent has different equilibrium positions for the different diabatic states.  The solvation interaction is described using a model of charge transfer between conducting spheres in a linear dielectric, with sphere radius and inter-sphere distances characteristic of the monomethine cyanine.  This model, which is quite well-established in the chemical physics literature related to electron-transfer reactions\cite{Nitzan2006marcus}, has also been used in a description of anomalous solvatochromism in charged push-pull polyenes (asymmetric polymethine cyanines)\cite{Laage2003}.  We have neglected the contribution of solvation effects to the coupling element, as has also been done in other studies of nonequilibrium solvation during photoisomerization reactions\cite{Malhado2011}.
\end{itemize}

\section{Results}\label{sec:results}
\subsection{General remarks and classification of PES}
	The potential energy surfaces given by the model can be qualitatively classified according to the "phase diagram" in Figure \ref{fig:phase_diagram}.  The figure classifies potential energy surfaces according to three yes/no questions.  These are:
\begin{itemize}
\item Will TICT states form spontanously after electronic excitation?
\item Is there a barrier to twisting on the excited state surface?
\item Do gapped TICT states exist?  A TICT state is gapped if it cannot be brought into coincidence with a conical intersection seam for any configuration of the solvent.
\end{itemize}  

	For parameters corresponding to typical monomethine cyanines in solvents of weak to moderate polarity, our model describes favorable (spontaneous) formation of twisted internal charge-transfer (TICT) states on the excited-state PES.  Twisting should be barrierless for dyes with electronic excitation energy above the far-red, but a barrier can exist for monomethines with weak electronic coupling or with weak $\pi$-bonding interactions.  Numerical estimates that we have performed for limiting cases suggest the barrier, if it exists, is probably small (a few $k_B T$, thermal energy at room temperature). 

	Our model can describe the differential stabilization of twisting channels by the solvation field in a monomethine, and can also describe movement of the conical intersection between the twisting channels in response to changes in solvent configuration.  This is an important aspect of the model, which is worth highlighting now.  This behavior is summarized in Figure \ref{fig:movingCI}.  This shows that our model, despite its simplicity, can describe the environmental selection of different twisting channels by an environmental field, such as has drawn interest in the literature on fluorescent proteins\cite{Olsen2010} and photoactive yellow proteins\cite{Ko2008}.  
	
	Our model suggests that, for typical monomethine cyanines with modest-sized endgroups (a single substituted heterocycle) in typical solvents (for which we expect $J\lessapprox \lambda$), there is no minimum on the excited state potential energy surface.  The bottom of the excited state potential corresponds to a conical intersection seam. This seam grows along the "valley floor" of the excited state PES.  For monomethine cyanines with modest-sized rings (containing one substituted ring, $r \sim 2\r A$), there are no gapped TICT states in solvents of even moderate polarity.  For monomethine cyanines in weak solvents and/or with large rings, our model predicts that gapped TICT states may exist.  If diffusion within the channels is slow enough, then these gapped TICT states may transiently accumulate excited-state population.  

\subsection{When are TICT states favorable? (Blue line in Fig. \ref{fig:phase_diagram})}
\begin{figure}
\includegraphics[width=7cm]{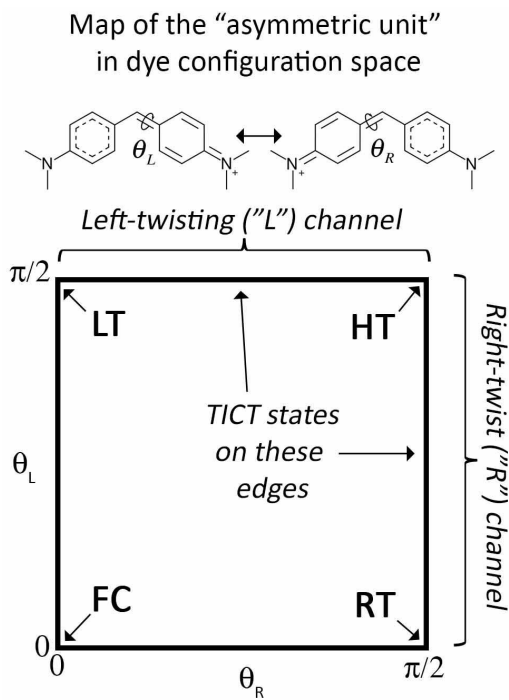}
\caption{\label{fig:atlas}A map of the "asymmetric unit" in dye configuration space, with labelling of landmarks referred to in the text.  The space, spanned by the twisting angles shown at top, is periodic and infinite.  In this paper, we focus solely on the patch shown, which is spanned by intervals of $[0,\frac{\pi}{2}]$ in both the twisting angles $\theta_L$ and $\theta_R$.  The corners of the patch are landmarks frequently mentioned by acronymn in the text: the Franck-Condon (\textbf{FC}) configuration, the left-twisted configuration (\textbf{LT}), the right-twisted configuration (\textbf{RT}) and a "hula-twist" (\textbf{HT}) configuration\cite{Liu1986} where both bonds are twisted.  Initial photoexcited population would be prepared at \textbf{FC} because the transition is bright there and it coincides with the ground state minimum.  The boundaries of the patch opposing \textbf{FC} and connecting the twisted configurations are where the twisted internal charge-transfer (TICT) configurations are found.  The TICT configurations are divided into the "L" TICT channel, consisting of configurations where $\theta_L=\frac{\pi}{2}$, and the "R" TICT channel $(\theta_R=\frac{\pi}{2})$.  Conical intersections between the adiabatic states in our model only occur in the TICT channels.  Our model doesn't distinguish concerted (even) and "hula" (odd) combinations of the torsions, nor does it distinguish between different \textit{cis}/\textit{trans} isomers of a given dye.}
\end{figure}

\begin{figure}
\centering
\includegraphics[width=10cm]{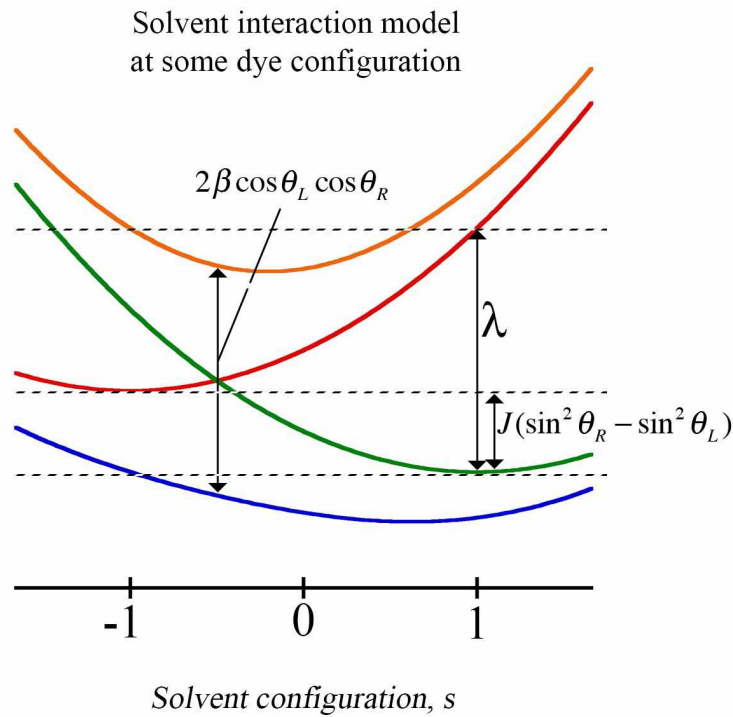}
\caption{\label{fig:Nitzan} 
A schematic diagram of the solvation interaction in our model.  Shown are the (quadratic) potential energy surfaces of the diabatic  and adiabatic states along a single solvation coordinate \textit{s}, which gauges the polarization of the solvent.  The solvent is in equilibrium with diabatic state $\lvert R \rangle$ (green line) at $s=1$ and in equilibrium with state $\lvert L \rangle$ (red line) at $s=-1$.  The diabatic gap is modulated by the torsion angles of the dye, $\theta_L$ and $\theta_R$ (cf. Figure \ref{fig:atlas}) through the torsion coupling constant $J$, which is the energy of twisting the $\pi-bond$ to the bridge.  The splitting of the adiabatic ground (blue) and excited (orange) states has contributions from the diabatic gap and the diabatic coupling matrix element. The diabatic coupling is modulated by the torsion angles through the electronic coupling constant $\beta$, and is equal to half of the electronic excitation energy at the ground state (planar) minimum.  At the solvent configuration where the diabatic states cross, the adiabatic states are split by an amount equal to twice the diabatic coupling.  The diabatic solvent reorganization energy $\lambda$ is the energy required to displace one diabatic state from its own equilibrium solvent configuration to the equilibrium configuration of the other.  It is the same for both diabatic states.
}
\end{figure}

\begin{figure}
\includegraphics[width=12cm]{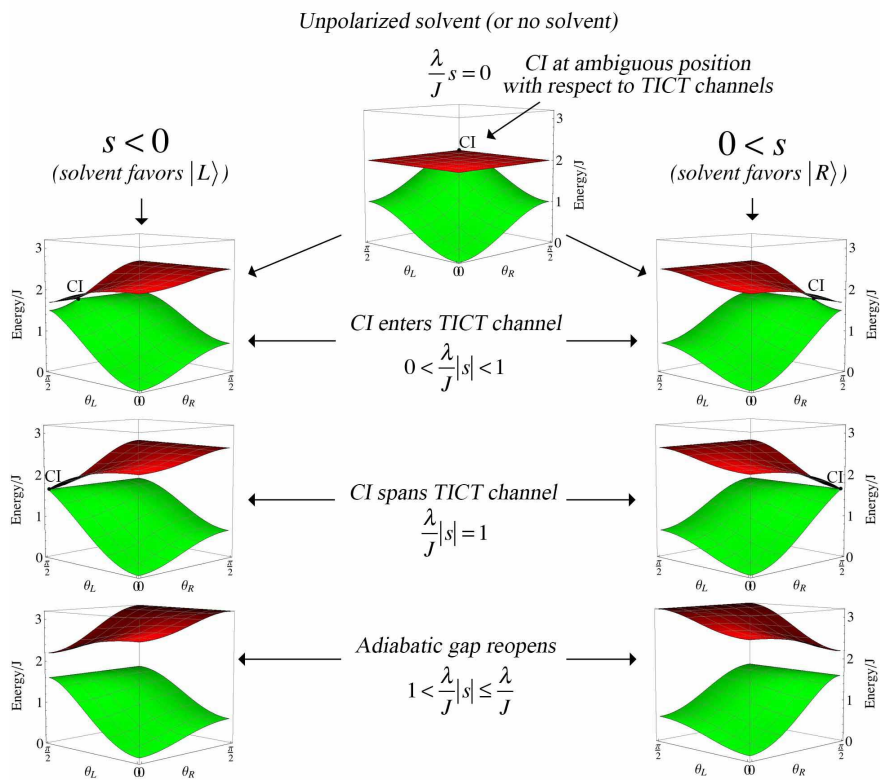}
\caption{\label{fig:movingCI}Changes in the potential surfaces of the dye in response to changes in the solvent configuration.  For an unpolarized solvent, such as at the ground state minimum, the excited state PES is ambiguous with respect to the L and R TICT channels (c.f. \ref{fig:atlas}), and the conical intersection (CI) between adiabatic states is positioned ambiguously, where both bonds are equally twisted.  As the solvent polarizes in one or the other direction, one channel is stabilized over another and the CI moves into the stabilized channel.  This corresponds to positioning the CI at dye configurations where one bond is twisted ($\theta=\frac{\pi}{2}$) and the other is less twisted ($\theta<\frac{\pi}{2}$).  For moderate polarization of the solvent, the CI is located in one of the twisting channels.  At a critical polarization $s=\pm\frac{J}{\lambda}$, the CI reaches the "end" of the TICT channel, where one bond is twisted and the other planar.  For polarizations beyond this, the adiabatic gap reappears and the excited state is destabilized. The solvent coordinate is bound in the interval $[-1,1]$, so that not all regimes are accessible if $\lambda<J$. The parameters used here were chosen for Michler's hydrol blue ($\beta=1.0eV,J=1.0eV,r=2.3\r A, R=4.6\r A$) in acetonitrile ($\lambda=1.6eV$).}
\end{figure}

	The twisted internal charge-transfer configurations/states in our model occur where the coupling matrix element vanishes $-2\beta\cos\theta_L\cos\theta_R=0$.  This occurs on the lines $\theta_L=\frac{\pi}{2}$ and $\theta_R=\frac{\pi}{2}$, limiting focus to the range indicated in \ref{fig:atlas}. We will divide the TICT states by dividing them into two "channels": the R channel ($\theta_R=\frac{\pi}{2}$) and the L channel ($\theta_L=\frac{\pi}{2}$).  The points corresponding to these channels are also indicated in Figure \ref{fig:atlas}.	

	TICT states will only form spontaneously on a relevant timescale if they are energetically favorable. For a single coordinate, this corresponds to either of the two left panels of Figure \ref{fig:TICTOutcomes}. The excited-state energy (relative to \textbf{FC}) $\Delta E_1(\theta_L,\theta_R,s)=E_1(\theta_L,\theta_R,s)-E_1(0,0,0)$ is defined in the $R$ and $L$ channels as
  
\begin{align}\label{eq:E1def}
\Delta_R E_1\left(\theta_L,s\right)=\frac{J}{2}\left(\sin^2\theta_L+1\right)
+\lambda\left(\frac{s}{2}\right)^2 +\frac{1}{2} |\lambda s - J \cos^2\theta_L|-\beta\\
\Delta_L E_1\left(\theta_R,s\right)=\frac{J}{2}\left(\sin^2\theta_R+1\right)
+\lambda\left(\frac{s}{2}\right)^2 +\frac{1}{2} |\lambda s + J \cos^2\theta_R|-\beta
\end{align}

	The absolute value function introduces a derivative discontinuity where it's argument is zero.  In this case, the argument is the adiabatic gap, and the derivative discontinuity gives the position of a conical intersection within the TICT channels.  

	The excited-state energy in the $L$ channel can be written analytically on either side of the conical intersection as

\begin{align}\label{eq:L_channel}
\Delta_L E_1\left(\theta_R,s\right)=J+\lambda \left(\left(\frac{s}{2}\right)^2+\frac{s}{2}\right)-\beta,\quad  0\leq J\cos^2\theta_R+\lambda s \\
\Delta_L E_1\left(\theta_R,s\right)=J\sin^2\theta_R+\lambda \left(\left(\frac{s}{2}\right)^2-\frac{s}{2}\right)-\beta,\quad  J\cos^2\theta_R + \lambda s \leq 0
\end{align}

and in the $R$ channel as

\begin{align}
\label{eq:R_channel}
\Delta_R E_1\left(\theta_L,s\right)=J\sin^2\theta_L + \lambda \left(\left(\frac{s}{2}\right)^2+\frac{s}{2}\right)-\beta, \quad 0\leq \lambda s-J\cos^2\theta_R \\
\Delta_R E_1\left(\theta_L,s\right)=J+\lambda \left(\left(\frac{s}{2}\right)^2-\frac{s}{2}\right)-\beta, \quad -J\cos^2\theta_L + \lambda s \leq 0
\end{align}.

\begin{figure}
\includegraphics[width=10cm]{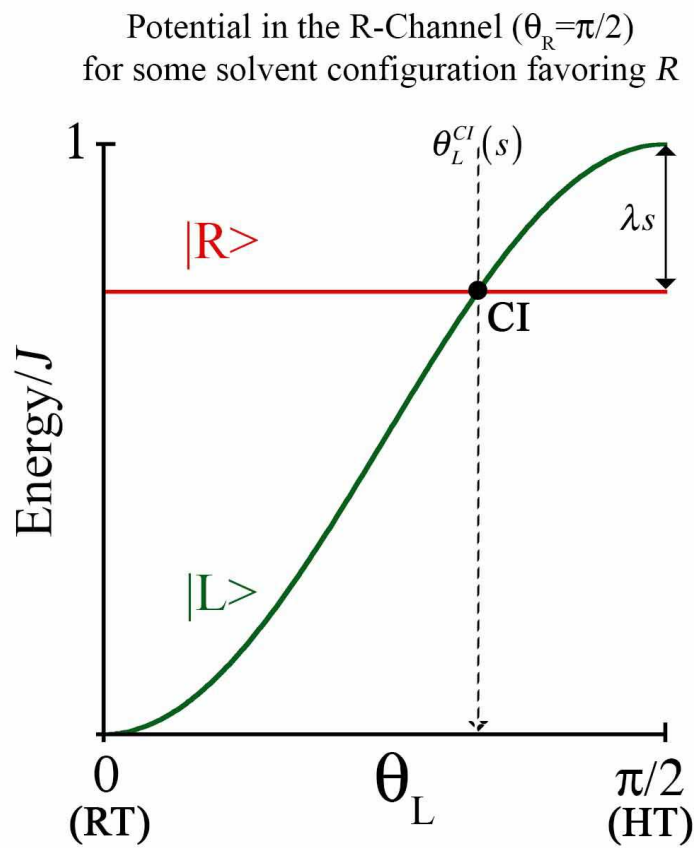}
\caption{\label{fig:Rchannel}Movement of the conical intersection in the TICT channels can be understood by examining the diabatic states within a single channel (here the R channel, for which $\theta_R=\frac{\pi}{2}$, for a solvent configuration favoring $\lvert R \rangle$).  \textit{The diabatic coupling vanishes everywhere in the figure}, so the adiabatic states are given by taking the lowest and highest value at every point, and the crossing point is a conical intersection between adiabatic states.  The potential for diabatic state $\lvert R \rangle$ in the channel is flat, because the $\pi$-bond in this state is already twisted.  The potential for $\lvert L \rangle$ rises as $\theta_L \rightarrow \frac{\pi}{2}$.  For an unpolarized solvent (such as at the ground state minimum), the CI is located at HT, a configuration where both bonds are twisted.  As the solvent moves to stabilize $\lvert L \rangle$, the excited state energy at LT decreases, and the CI moves from HT towards RT (to the left of the figure).  When the solvent stabilization $\lambda s=J$, the CI has reached RT.  For larger polarizations of the solvent, the $\pi$-bond stabilization energy can no longer compete with the energy of the solvent interaction.  Then, the adiabatic gap reopens and the CI disappears.}
\end{figure}

	The situation is described visually in Figure \ref{fig:Rchannel}, which shows the energies in the $R$ channel for some solvent configuration that puts the CI in that channel.  Since the coupling vanishes in the TICT channels, the diabatic crossing point is also a conical intersection between adiabatic states.  We can see immediately that diabatic state $\lvert R \rangle$ favors unwinding the angle $\theta_L$, while $\lvert L \rangle$ is invariant to $\theta_L$.  The coupling vanishes in the channel, so the lowest adiabatic excited state energy in the channel can never be lower than the adiabatic excited state energy at the geometry \textbf{RT}, corresponding to $\theta_L=0$.

	It follows that the lowest excited-state energy (relative to \textbf{FC}) in the $R$ channel equals the excited-state energy at \textbf{RT}.  Analogous reasoning says that the lowest energy in the $L$ channel is the energy at \textbf{LT}.  The energies at these points is written on either side of the conical intersection as

\begin{align}\label{eq:LTenergy}
\Delta_{LT} E_1\left(s\right)=J+\lambda \left(\left(\frac{s}{2}\right)^2+\frac{s}{2}\right)-\beta, \quad 0\leq \lambda s -J \\
\Delta_{LT} E_1\left(s\right)=\lambda \left(\left(\frac{s}{2}\right)^2-\frac{s}{2}\right)-\beta, \quad \lambda s-J\leq0
\end{align},

while the excited-state energy at \textbf{RT}, which similarly bounds the energy in the $R$ channel from below, we have

\begin{align}\label{eq:RTenergy}
\Delta_{RT} E_1\left(s\right)=\lambda \left(\left(\frac{s}{2}\right)^2+\frac{s}{2}\right)-\beta, \quad 0\leq \lambda s -J \\
\Delta_{RT} E_1\left(s\right)=J+\lambda \left(\left(\frac{s}{2}\right)^2-\frac{s}{2}\right)-\beta, \quad \lambda s-J\leq0
\end{align}.

	We can see that the lowest energy at \textbf{RT} is achieved for $\lambda<J$ at $s=1$, whereas for $J\geq\lambda$ it is achieved at $s=\frac{J}{\lambda}$, which is at the conical intersection.

	It follows that the lowest energy that is achievable in the TICT channels is the energy at \textbf{RT} (or \textbf{LT}) \textit{at the solvent configuration which brings the conical intersection to that point}.  This energy is given (relative to \textbf{FC}) by

\begin{equation}\label{eq:E1min}
\Delta E_1^{min}=J-\beta-\frac{\lambda}{4}
\end{equation}.

	This equation gives the lowest (most favorable) excited-state energy that can be obtained for any TICT state in our model (relative to \textbf{FC}).  If it is negative, then there exist TICT states that are energetically accessible from \textbf{FC}.  If it is positive, then the TICT states are not accessible from \textbf{FC}.  In the former case, the TICT states are expected to form spontaneously in the excited state, on a timescale that will depend on the intervening potential (cf. Figure \ref{fig:TICTOutcomes}, difference between two left-hand panels and next section).  In the latter case, the molecule will remain on the excited state in a region of strong radiative coupling and would be expected to decay by fluorescence.  If the energy at \textbf{RT}/\textbf{LT} vanishes, then there is no driving force for twisting.  In the case of a flat, barrierless surface, Equation \ref{eq:trot} is still useful if the average torque over a radian is taken as thermal energy at room temperature, $k_B T$.  In that case, for a rotor with $r \sim 2 \r A$ in a medium with viscosity 1 cP, the resulting timescale is still in the picosecond range.  Non-radiative decay should still win over fluorescence, even if the potential is completely flat (for typical solutions at room temperature).

\subsection{Is there a barrier? (Green line in Fig. \ref{fig:phase_diagram})}	
	The excited-state twisting process in monomethine cyanines has been invoked as an example of a barrierless viscosity-controlled process.\cite{Aberg1994}  Quantum chemistry calculations on monomethine cyanines and similar systems usually indicate no barrier or small barrier.\cite{Improta2005}\cite{Olsen2008}\cite{Weigel2012} The accepted limits of quantum chemistry estimates of excited-state energy differences are of the order of a few $k_B T_r$ (even where the Born-Oppenheimer approximation is appropriate), so predictions of a barrier smaller than this are not conclusive.\cite{Azizi2006}

	Here, we examine what can be expected, with respect to the presence or absence of a barrier to twisting, on the basis of the few enumerable assumptions included in our model.

	The question of whether or not there is a barrier to twisting boils down to the sign of the eigenvalues of the second-derivative matrix (Hessian) with respect to the twisting displacements at \textbf{FC}.  The solvent coordinate is always bound in the adiabatic excited-state of our model (as can be verified by consideration of Figure \ref{fig:Nitzan} and remembering that $\lambda\geq 0$).  The Hessian is diagonal at \textbf{FC}, and the force constants associated with the two twisting displacements are equal by symmetry.  The excited-state force constants with respect to these displacements have the particularly simple form

\begin{equation}\label{eq:dble_derivativ}
\frac{\partial^2 E_1(0,0,0)}{\partial \theta_R^2}=\frac{
\partial^2 E_1(0,0,0)}{\partial \theta_L^2}=J-\beta,
\end{equation}.

\begin{figure}
\centering
\includegraphics[width=13cm]{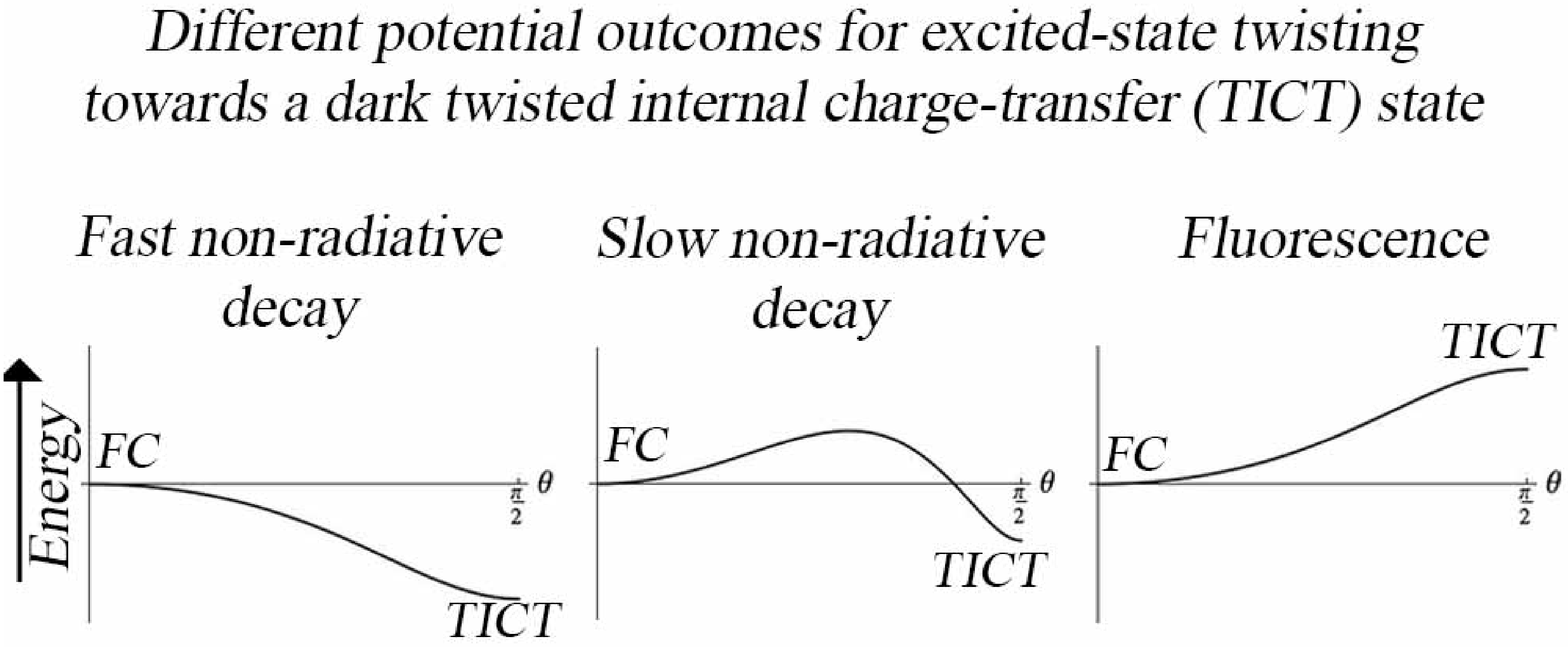}
\caption{\label{fig:TICTOutcomes}Three distinct possible outcomes for an excited-state twisting reaction leading to a dark twisted internal charge-transfer (TICT) state.  The reaction coordinate is taken to be a single bond-twisting angle $\theta$, which connects the Franck-Condon (FC) region on the excited-state potential energy surface with the TICT state at $\theta=\frac{\pi}{2}$.  \textit{(left)} If the reaction is downhill and barrierless, fast non-radiative decay is expected, with fast (fs-ps) decay of fluorescence in solutions of low (~1 cP) viscosity.  \textit{(middle)} If there is a barrier along the reaction coordinate, but the reaction is still favorable, then slower non-radiative decay is expected, with the rate depending on the barrier height.  For barriers less than a few thermal energy units $k_BT_r$, only transient fluorescence is expected; for larger barriers steady-state fluorescence may be observed. \textit{(right)} For an uphill reaction, the TICT states do not form spontaneously in significant amounts.  In this case, the dye will decay by emission of fluorescence.}
\end{figure}

	Negativity of the twisting force constants at \textbf{FC} implies accessibility of TICT states, as can be seen by comparing Equations \ref{eq:dble_derivativ} and \ref{eq:E1min}.  If $J-\beta \leq 0$, then the model predicts barrierless descent to the TICT manifold.  This would correspond to the case of the far left panel in Figure \ref{fig:TICTOutcomes}.  If $0 < J-\beta \leq \frac{\lambda}{4}$, then \textbf{FC} is bound but the TICT states are still favourable, and twisting proceeds with a barrier (middle panel of Figure \ref{fig:TICTOutcomes}).  If the TICT states are inaccessible ($\frac{\lambda}{4}<J-\beta$), then the molecule will decay by fluorescence (right panel, Figure \ref{fig:TICTOutcomes}).

	Since the adiabatic transition is dipole allowed at \textbf{FC}, but forbidden in the TICT channels, the transient fluorescence decay measures exit from the region of \textbf{FC}.  We expect the presence of the barrier to lengthen the fluorescence decay.  However, the time to internal conversion on the far side of the barrier should be independent of the barrier height, so that in the activated regime we expect other contributions to the ground-state recovery time to be unaffected.  The ground-state recovery time would be expected to be dominated by the fluorescence decay time in this case.

	Comparison of Equations \ref{eq:E1min} and \ref{eq:dble_derivativ} shows that $J\geq \beta$ implies that twisting is unfavorable in sufficiently non-polar solvents.  However, Table \ref{tab:solv_reorg_en} shows also that even for relatively non-polar solvents we still have $\lambda \geq 0.5eV$.  For Michler's hydrol blue (cf. Figure \ref{fig:Scheme}), we have the interesting case of $J\sim \beta$, so that the driving torque for twisting in gas phase is expected to vanish.  We have tested this assertion for Michler's hydrol blue using multi-reference perturbation theory estimates of the excited-state energy at the ground state minimum (a model of \textbf{FC}) and at a twisted excited-state minimum (a model of \textbf{RT} or \textbf{LT}) and found that this is true to within the expected accuracy of the calculations.  In this case, solvation in weakly polar solvents such as eg. chlorobenzene will induce a bias of $\sim 0.2eV$, which should be sufficient to quench fluorescence.

	The reddest dyes with absorbance in the visible would have $\beta \sim 0.8eV$.  We have numerically calculated the barrier height for a dye with $\beta=0.8eV$,$J=1.0eV$ with dimensions comparable to Michler's hydrol blue ($r=2\r A$, $R=4\r A$) in chlorobenzene ($\lambda=0.8eV$) and found it to be less than $3k_BT$.  Assuming an attempt frequency of $\sim 1ps$ for a torsional barrier, this would not be sufficient to engender steady-state fluorescence in the dye for moderately polar solvents $\lambda \sim 1eV$. 

	The observation of a barrier to twisting in our model must be approached with awareness that we are leaving out potentially important intramolecular degrees of freedom.  For example, our model does not consider bond vibration displacements, to which the states in these systems are known to be coupled.  Laage and Hynes have suggested a form for the diabatic coupling that is formally identical to the solvation model we use here.\cite{Laage2003}  It has been pointed out that the coupling of solvation and stretching displacements through the same electronic degree of freedom (the diabatic gap), can amplify the effects of either.\cite{Painelli1999}  If a solvation coordinate were included in this way, with a reorganization energy of order $\sim0.1eV$,  it would act to reduce the barrier height and increase the driving force for reaction.  This may lower the barrier to the point that it becomes irrelevant for typical monomethine cyanines.

	We conclude that our model predicts barrierless or near-barrierless excited state twisting for typical solvated monomethine cyanines, but we are unable to rule out the possibility that a monomethine cyanine may exist, if the dye has near-IR absorbance and large $\pi$-bonding interactions, that would show steady-state fluorescence in sufficiently non-polar solvents.
 
\begin{figure}
\includegraphics[width=12cm]{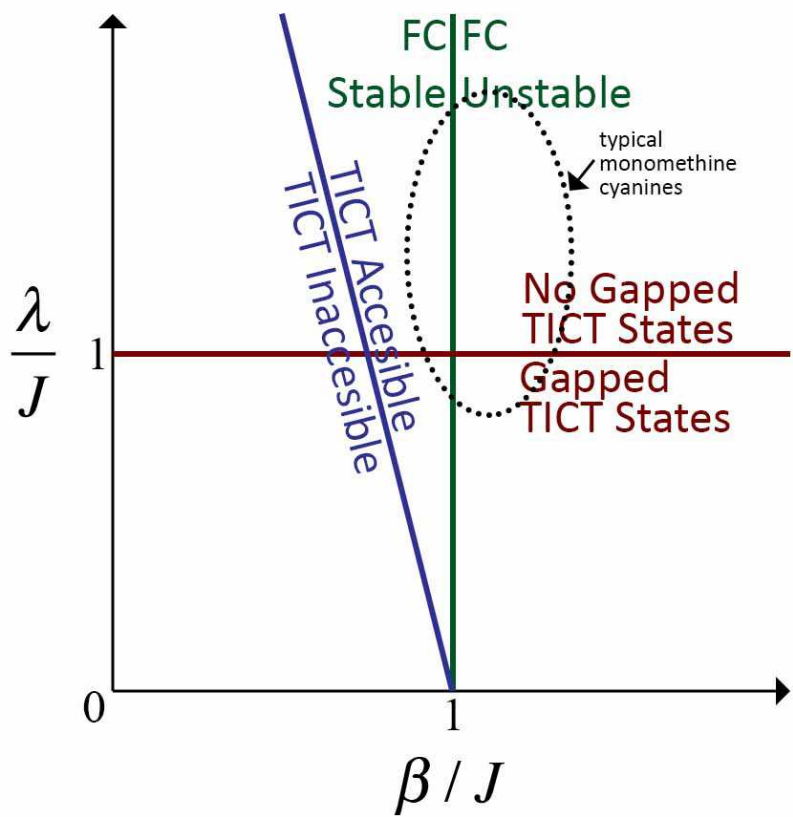}
\caption{\label{fig:phase_diagram}This "phase diagram" summarizes the qualitative features of potential energy surfaces (PESs) provided by the model in distinct regimes of the parameters $J,\lambda,\beta$.  The parameter regimes characteristic of typical monomethine cyanines are inside the dotted ellipse.  PES given by the model are classified according to the disposition with respect to three lines drawn.  Above the blue line, spontanous formation of dark TICT states is expected.  Below the line, the dye will decay by fluorescence.  To the left of the green line, the (planar) \textbf{FC} configuration is stable (left) on the exited-state PES.  If TICT states are still favorable, then there is a barrier to reach them.  To the right of the line the planar (FC) state is unstable on the excited state surface, and twisting occurs without barrier.  Below the red line, there are TICT states that are adiabatically gapped for all solvent configurations.  Above the red line, the TICT states are ungapped, in the sense that the CI can be made to coincide with any TICT geometry for some configuration of the solvent.}
\end{figure}

\subsection{Do gapped TICT states exist? (Red line in Fig. \ref{fig:phase_diagram}).}
	We call at particular TICT state (geometry) \textit{gapped} if it cannot be made to coincide with the conical intersection seam for any configuration of the solvent.
		
	It has been suggested that the timescales observed in ultrafast spectroscopic experiments on monomethine cyanines may be interpreted on the basis of a partitioning of the dynamics between (fast) descent into a basin corresponding to a twisted configuration, followed by diffusion to the sink or conical intersection.\cite{Dietzek2009}  This notion presupposes the existence of an adiabatic twisted minimum on the excited state surface, in which diffusion could occur.  In this section, we address the question of whether a true (i.e. gapped) excited state minimum can exist in the TICT manifold.

	To answer the question, we analyze the extent and location of conical intersections in the model.  The conditions for intersection of adiabatic states is that the diabatic gap and coupling vanish simultaneously.  This gives

\begin{equation}\label{eq:CI_conditions1}
-2\beta\cos \theta_L\cos \theta_R=0\\
\end{equation}

for the coupling, and
\begin{equation}\label{eq:CI_conditions2}
J\left(\sin^2\theta_L-\sin^2\theta_R\right)-\lambda s = 0.
\end{equation}
for the gap. 

	Equation \ref{eq:CI_conditions1} is only (and always) satisfied in the TICT channels, so that conical intersections will only ever occur in the channels.

	Equation \ref{eq:CI_conditions2} can be used to derive the solvent configuration that brings the conical intersection into coincidence with any point in the TICT channels.  This is, for the $L$ and $R$ channels,

\begin{equation}\label{eq:seam_range}
s^{CI}(\theta_L,\theta_R)=\frac{J}{\lambda}\left\lbrace
\begin{array}{cc}
\cos^2\theta_R,&|\theta_L|=\frac{\pi}{2}\\
-\cos^2\theta_L,&|\theta_R|=\frac{\pi}{2}
\end{array}
\right. 
\end{equation}

	This shows that the CI will be in the $R$ channel for $s\ge 0$ and the $L$ channel for $s \le 0$.   Again, we have concentrated solely on the "patch" shown in Figure \ref{fig:atlas}; a more general statement would include periodic images of the CI.

	It is clear that \ref{eq:CI_conditions2} can only be solved for $s \leq \frac{J}{\lambda}$ if $J\leq \lambda$, since the range of $s$ is limited to the interval $[-1,1]$.  For solvent configurations $s \geq \frac{J}{\lambda}$, no conical intersection exists (cf. Figure \ref{fig:movingCI}, bottom row).  

	For any set of model parameters $J$, $\beta$, $\lambda$ such that $\lambda \leq J$, the range of motion of the CI into the channels is limited to angles $\theta^{CI}_{min} \leq \theta_{L,R} \leq \frac{\pi}{2}$.  Accordingly, there is an adiabatic gap remaining at angles $\theta_{L,R}$ less than some minimum angle $\theta^{CI}_{min}$ for which the CI may exist.  If TICT states exist that are gapped for all solvent configurations, then these may accumulate population on the excited state surface, which will decay following diffusion to the vicinity of the CI seam.

	If the adiabatic excited state PES does possess a minimum, and if this basin is sufficiently removed in energy and configuration space from the intersection seam such that transient steady-state population accumulates there, we would expect the ground state recovery time to increase without consummate lengthening of the fluroescence decay time.  This is because the TICT states are dark, so accumulated population in the TICT channels will not contribute to the fluorescence.

	Table \ref{tab:solv_reorg_en} says that, for a dye with dimensions similar to Michler's hydrol blue ($r\sim 2-3\r A$ , $R\sim 4-7\r A$), the reorganization energy should be of order 1 eV even for weakly polar solvents.  Accordingly, we expect that $\theta^{CI}_{min} \sim 1$ for these solvents.  This implies that the conical intersection can access most of the TICT configurations in the channels.  However, even for an angular interval of $\sim 0.1$ radian, we expect that the diffusion time in a medium of viscosity $~1$ cP might be $~10-100ps$, and so this should be detectable as a delay in the ground state recovery time that is not apparent in the fluorescence decay time.  

	For dye-solvent systems with $\lambda\geq 1eV$, the CI can be brought into coincidence with any point in the TICT channels.  Since the TICT channels represent the bottom of the excited-state potential (c.f. discussion relevant to blue line in Figure \ref{fig:phase_diagram}, above), this implies that a CI spans the bottom of the excited state potential of the model in this regime.  This regime is the one that we expect to be relevant to most spectroscopic experiments using monomethine cyanines in moderate- to strongly polar solvents (such as eg. small-chain alcohols).

\begin{figure}
\includegraphics[width=15cm]{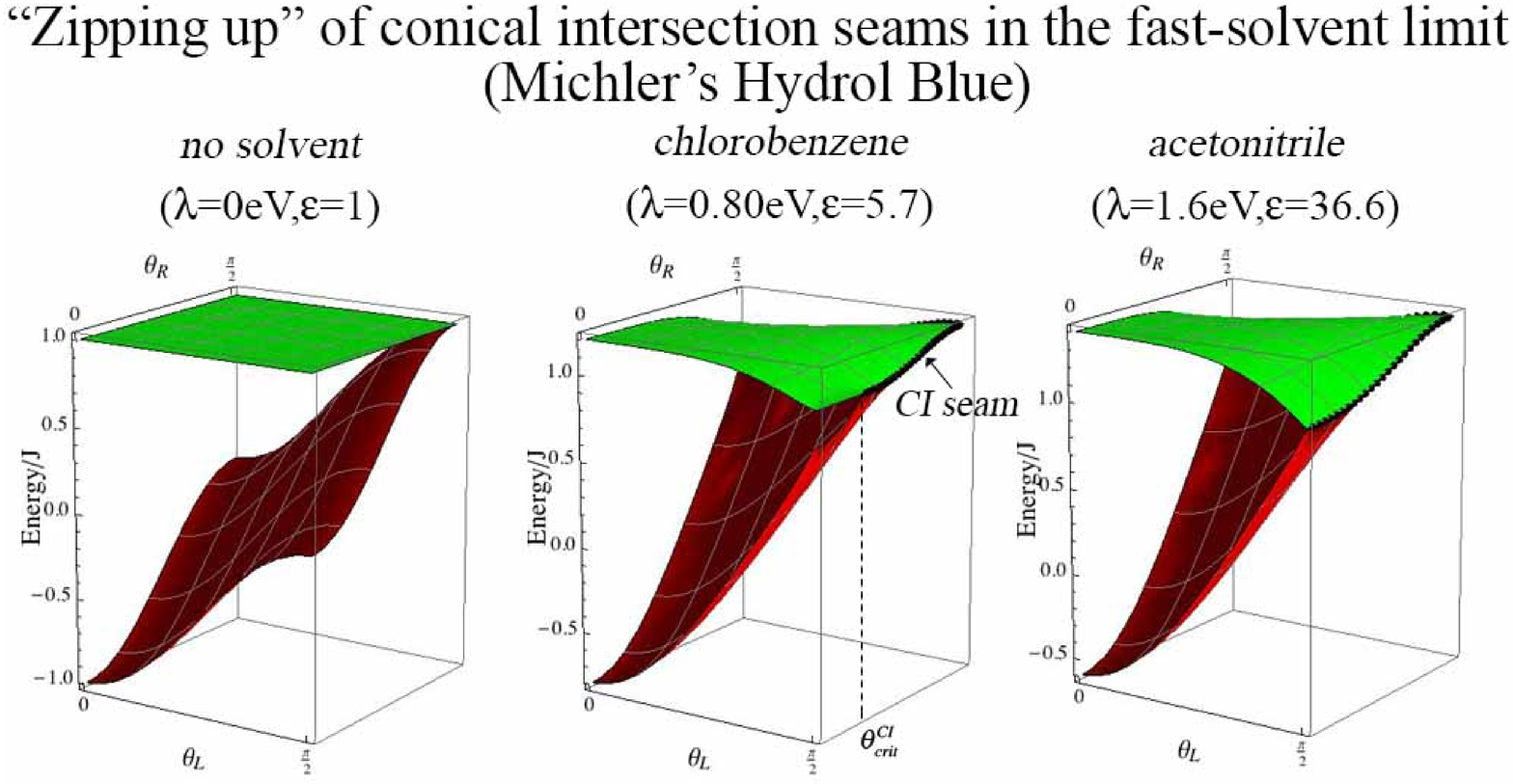}
\caption{\label{fig:zippingupci}Potential energy surfaces for Michler's Hydrol blue ($\beta=2.0$ eV, $J=1.0$ eV) in the fast-solvent limit, in gas phase (left), chlorobenzene (middle) and acetonitrile (right).  Reorganization energies and static dielectric constants for each solvent are shown.   In the fast solvent limit, growth of the CI seam along the bottom of the excited state potential energy surface (PES) appears as a "zipping up" of the CI seam along the twisted internal charge-transfer (TICT) states.  This is because, for any TICT configuration, the solvent configuration that yields the lowest excited-state energy is the one that brings the CI to that point.  The fast-solvent limit is taken at any dye configuration by letting the solvent equilibrate to the excited state of the dye at that configuration.  Shown are fast-solvent-limit PES for Michler's hydrol blue ($J=1.0$ eV, $\beta=1.0$ eV, $r=2.3$ $\r A$, $R=4.6$ $\r A$) in gas phase (left), a weakly polar solvent (chlorobenzene, middle) and a strongly polar solvent (acetonitrile, right).  Even for chlorobenzene, the CI seam spans most of the TICT configurations.  The PES shown for acetonitrile is identical in any solvent for which $\lambda\geq1$, since at that point the TICT states are already "all zipped up" ($J=1$ eV for this case).  The fast-solvent limit is not physically realistic for these systems, but provides good illustration of the growth of the CI seam in solvents with increasing susceptibility.
}
\end{figure}

	The growth of access to the TICT channels by the CI can be dramatically visualized by taking the fast-solvent limit, where the solvent is allowed to equilibrate to the dye at every fixed dye configuration.  This limit yields two-dimensional potential energy surface such as shown in Figure \ref{fig:zippingupci} for Michler's hydrol blue.  In the fast-solvent limit, the coincidence of the CI with the bottom of the excited-state PES is apparent as a "zipping up" of the TICT states as the seam extends into the channels.  The results shown in Figure \ref{fig:zippingupci} show that even for weakly polar solvents, the intersection can already extends over most of the TICT configurations.  The PES given for acetonitrile in the figure is the same for \textit{all} solvents with $\lambda \geq J$.  This is because solvent configurations that are sufficiently polarized to re-introduce an excited-state gap also raise the excited-state energy, so that the force on the solvent coordinate at these configurations will drive the system back towards the intersection.  Accordingly, these solvent configurations are never sampled in the fast-solvent limit.
 
	The phenomenon of "zipping up" of intersection seams in the fast-solvent limit has also been described by Burgardt and co-workers for a model of the photoisomerization of a protonated Schiff base\cite{Burghardt2004}.  This phenomenon can be traced in both cases to the diagonal form of the solvation coupling used in the Hamiltonian Equation \ref{eq:Ham_mat}. 

	Erez et al. have showed for the molecular rotors Thioflavin-T and Auramine-O, that the nonradiative decay timescale is 1-2 orders of magnitude slower than the transverse dielectric relaxation timescale in a variety of glass-forming solvents\cite{Erez2012d}.  This is consistent with a twisting timescale that is of the same order as the \textit{longitudinal} Debye relaxation time.  The longitudinal time will an order of magnitude lower than the transverse time in lower alcohols.  The longitudinal relaxation time is the relevant time for solvation following a change in the charge distribution\cite{Nitzan2006sol}.  These points emphasize that only qualitative insights can be obtained from considering the fast-solvent limit, which is not believably achieved in real experiments on symmetric monomethines.  

\section{Discussion}
	We have proposed a valence-bond model of non-equilibrium solvation effects on the excited-state twisting reaction of a general monomethine cyanine dye. The model has been made intentionally simple, so as to explore the consequences of a few specific ideas.  

	The solvation model is similar to a previously published model of anomalous solvatochromism charged push-pull polyenes\cite{Laage2003}. It differs by the introduction of potentials for twisting distinct $\pi$-bonds in the diabatic states, and the introduction of a twist-dependent coupling that goes as product of $\pi$ orbital overlaps across the bridge.  The functional forms used can be justified using valence-bond models\cite{Shaik2004}.  

	Our results suggest that the concept resonance in dyes, traditionally used to describe optical properties\cite{Platt1956,Platt1961,Brooker1942}\cite{Moffitt1950,Lu1994,Painelli1999,Thompson1998,Thompson1999}, may also explain non-radiative decay.  Since the resonance notation of Figure \ref{fig:Scheme} describes a transfer between states with different charge- and bond-order localization, the location of the barrier is different in the diabatic states.  The correlation of charge and $\pi$-bond order localization has important consequences for the structure of the excited state PES in the model, and the response to non-equilibrium solvation.  The assumption of a coupling that varies as a product of overlaps across the bridge bonds is a consequence of the negligibility of the direct (single particle) inter-ring charge-transfer matrix element, which is traditionally assumed.\cite{Pauling1939}

	The model broadly reproduces features of the ground and excited state PES that are observed in more complicated electronic structure calculations of monomethine cyanines\cite{Weigel2012}\cite{Improta2005} and related systems\cite{Olsen2010}\cite{Olsen2012a}\cite{BoggioPasqua2009}\cite{Knoch2014}\cite{Ko2008}.  These studies indicate that there are two energetically accessible twisting channels on the excited state surface, and that these channels are distinguished by the polarity of the adiabatic TICT transitions within the channels\cite{Improta2005}\cite{Weigel2012}, such that control of the pathway yield by environmental interactions is possible\cite{Knoch2014}\cite{BoggioPasqua2009}\cite{Ko2008}.  

	For parameter regions characteristic of monomethine dyes (c.f. Figure \ref{fig:Scheme}), there is a single charge-transfer conical intersection in the model that moves in response to changes in solvent configuration.  For an unsolvated dye or a dye in an unpolarized solvent, the intersection occurs at a geometry of the dye where both bonds are twisted.  This is an ambiguous position with respect to the two TICT channels.  It corresponds to either a concerted- or "hula-" twisted geometry\cite{Liu1986}.  As the solvent configuration changes, so as to solvate better one or the other diabatic state, the intersection moves into one of the TICT channels.  This means that the intersection will occur at a geometry where one bond is twisted more than the other.

	In general, \textit{ab initio} calculations on methine dyes show that minimal-energy conical intersections in symmetric monomethine cyanines occur at geometries where both bonds are twisted, but where the twist distribution is not symmetric\cite{Improta2005}.  This is also true in non-symmetric methines with very small diabatic gap\cite{Olsen2009}.  The reason why our model predicts a symmetric twist distribution can be traced to the neglect of other intramolecular displacements that could break symmetry in the any real specific case.   In any case, our results suggest that intersections in methines can move in response to solvation, and will move to the bottom of the excited-state PES for solvents of moderate polarity.  Decay via different twisting channels is favoured for distinct configurations of the solvent.  Migration of conical intersections towards the bottom of the excited state potential in solvated organics was observed in early QM/MM wavepacket simulations of solvated photoisomerization in methines\cite{Toniolo2004}.  Our model gives a simple and transparent description of this effect for the case of monomethine cyanines.

	Our model predicts that configurations with mixed twist distributions (for example, the "hula-twist" configuration \textbf{HT}) have relatively low energy on the excited-state PES.  In particular, these geometries are accessible from \textbf{FC} in gas phase.  This is contraindicated by \textit{ab initio} calculations on monomethine cyanines\cite{Improta2005}, as well as non-symmetric monomethines with small diabatic gap\cite{Olsen2009}.  The physical origin of the discrepancy is that, in our model, there is no additional penalty for breaking both $\pi$ bonds at once.  This could be added to our model easily, but would not qualitatively alter our main results.  
	
	We have provided a simple expression for the lowest energy possible in the TICT channels, relative to \textbf{FC}, that can occur in our model.  This is Equation \ref{eq:E1min}.  This can be used in conjuction with the timescale formula Equation \ref{eq:trot} to yield a timescale for rotation of one of the rotors in a monomethine.  An estimate of the relevant $torque$ in this case is the lowest energy in the channel (Equation \ref{eq:E1min}) divided by $\frac{\pi}{2}$ radians.  This yields the modified timescale formula

\begin{equation}\label{eq:trot2}
\tau_{TICT}=\frac{\pi \eta V}{\Delta E^{min}_1}
\end{equation}.

From this timescale, assuming no other non-radiative decay channels and also accounting for the existence of two rotors in monomethine cyanines, we can obtain an estimate of the fluorescence quantum yield $\Phi_f$ as

\begin{equation}\label{eq:QYield}
\Phi_f = \frac{1}{\frac{2 \tau_{rad}}{\tau_{TICT}}+1}
\end{equation}.

	For Michler's Hydrol Blue ($\beta = 1$ eV, $J = 1$ eV, $r = 2 \r A, R = 4 \r A$, $V = 34 \r A^3$) in methanol ($\eta = 0.6$ cP), dichloromethane ($\eta = 0.45$ cP) and chlorobenzene ($\eta = 0.8$ cP), this approach yields quantum yield estimates of $2 * 10^{-4}$, $2 * 10^{-4}$ and $7 * 10^{-4}$, respectively, assuming a radiative timescale of $\tau_{rad} = 2$ ns.  These estimates are in reasonable agreement with measured values of $2 * 10^{-4}$, $3 * 10^{-4}$ and $5 * 10^{-4}$, respectively.\cite{Momicchioli1993}
	
	Equations \ref{eq:QYield}, \ref{eq:trot2} and \ref{eq:E1min} together suggest that the fluorescence quantum yield of a monomethine cyanine is limited by the relative value of the driving torque in the excited state and the hydrodynamic friction.  For a given dye chromophore (i.e. given $\beta$, $J$), the driving torque will be modulated by the reorganization energy as Equation \ref{eq:E1min}, so that decreasing the reorganization energy of the environment of the dye will, in the absence of other considerations, give greater fluorescence, as will increasing the viscosity.  Alternatively, for a given medium, the driving force will be increased by reducing the electronic coupling or by increasing the hydrodynamic volume.  
	
	Assuming a $\pi$-bond stabilization energy of $J = 1$ eV, as we have here, implies that only monomethines with near-IR absorbance will have inaccessible TICT states.  Polar solvation can only accelerate TICT formation (in one channel or another).  In a structured environment, such as provided by a protein, it is possible that the elastic properties of the protein may provide an additional potential barrier, but there seems little scope for engendering fluorescence in fluid solution by the manipulation of the driving torque.  
	
	On the other hand, if the rotation timescale is linear in the viscosity as suggested by Equation \ref{eq:trot}, and given that the timescale for diffusion on a flat potential of $\sim 1$ ps for e.g. Michler's Hydrol Blue, one expects that a 1000-fold increase in the viscosity should be sufficient to achieve significant fluorescence.  This is consistent with changes in the fluorescence of Michler's Hydrol Blue recorded in a series of water-glycerol mixtures sampling this viscosity range.\cite{Momicchioli1993}  It is also broadly consistent with the radiative lifetime of Auramine-O and Thioflavin-T over a broader range.\cite{Erez2012d}
	
	Both the driving torque and the hydrodynamic friction depend on the spatial extent of the endgroups.  The driving torque varies inversely with $r$ (cf. Equation \ref{eq:reorg_e}), while the hydrodynamic friction scales with the volume (i.e. with $r^3$).  Accordingly, varying the spatial extent of the endgroups with bulky substituents may be an effective way to tune the solution fluorescence quantum yield of monomethine cyanines.
	
		Symmetric monomethines have been used as experimental models for the study of barrierless viscosity-controlled reactions\cite{Aberg1994,Dietzek2007a,Dietzek2007,Dietzek2009,VanderMeer2000}.  For the cyanine 1122C, Yartsev and coworkers have cited evidence of a thermalized excited state population\cite{Dietzek2009}.  Specifically, a fast (~1ps) component in the dynamics has been attributed to twisting motion towards an excited state basin, followed by depletion from the basin through the sink on the longer (~10ps) timescale\cite{Dietzek2009}.  For a dye such as 1122C, with rings about double that of Michler's Hydrol Blue, the reorganization energy in alcohol solution may be sufficiently low as to allow the existence of gapped TICT states, which could support adiabatic excited-state population diffusion.  For a dye of this size in methanol, we have estimated the timescale for TICT formation (as above, but with $r = 4\r A$, $R = 8 \r A$), and found it to be several picoseconds.  However, the timescale for diffusion on a flat potential (such as would be expected for population diffusing on the adiabatic excited state in a gapped region of the TICT channels) is several \textit{tens} of ps.  For dyes in moderately polar solution, for which the driving torque will be $\gtrapprox 10 k_B T$, the timescale for TICT formation may be distinctly faster than the timescale for flat potential diffusion of an adiabatically excited TICT population.  This suggests that our model may describe the transient accumulation of adiabatically excited TICT population for dyes with rings comparable to those in 1122C.  Further elaboration of this possibility would seem to require time-dependent simulations, and is beyond the scope of this paper.

\section{Conclusion}
	We have proposed and analysed a valence-bond model of non-equilibrium solvation in a monomethine cyanine dye.  The lowest adiabatic transition in these systems is traditionally described using a concept of resonance between quantum states with distinct charge and bond localization.  This results in strong coupling to the twist degrees of freedom and a twist-dependent charge-transfer character of the transition, all of which our model captures.  It also describes several features of the potential energy surfaces that have been observed with more complicated electronic structure models\cite{Improta2005}, including the occurrence of multiple low-energy TICT channels on the excited state surface, correlation between twisting and charge-transfer polarization of the transition in the channels, and the existence of low-energy conical intersections in the TICT channels.  We believe that our model may be the simplest that captures these three effects.
	
	Our modelling suggests that solvent effects are significant on the potential energy surfaces that govern photoexcited twisting in monomethine cyanines.  Polar solvation generally stabilizes TICT states, and can increase the driving forces for TICT formation.  We have provided a simple expression for the lowest possible energy (relative to \textbf{FC}) that can exist in the TICT channels of our model (Equation \ref{eq:E1min}).
	
	Our modelling suggests that TICT formation will be favorable for typical monomethine cyanines in solvents sufficiently polar to allow good solvation.  TICT formation should be barrierless for dyes with absorbance above ~600nm (for which $\beta \geq J \approx 1.0$ eV).  For dyes with absorbance at lower photon energy, our modelling predicts a small barrier may exist on the excited-state surface.  This barrier should be less than a few $k_BT$ for dyes with visible absorbance.  If a barrier exists, then our model also predicts that excited-state twisting will be unfavorable for the same dye in vacuum.
	
	A conical intersection grows along the bottom of the excited-state surface in solvents with increasing relative dielectric.  For monomethine cyanines with modest-sized rings in moderately polar solvents, our modelling predicts that no adiabatically excited TICT state can exist, as the bottom of the excited-state PES is completely spanned by a conical intersection seam.


%
%

%

\begin{acknowledgements}
This work was supported by Australian Research Council
Discovery Project DP110101580.
We thank A. Painelli, J.T. Hynes and T.J. Mart\' inez for helpful discussions.
\end{acknowledgements}

\providecommand{\latin}[1]{#1}
\providecommand*\mcitethebibliography{\thebibliography}
\csname @ifundefined\endcsname{endmcitethebibliography}
  {\let\endmcitethebibliography\endthebibliography}{}


\end{document}